\documentclass[doublecol]{epl2} 
\usepackage{amsmath}
\usepackage{graphics}
\usepackage{pdfpages}
\usepackage{amsfonts}
\usepackage{caption}
\usepackage{subcaption}
\usepackage{xcolor}
\usepackage{pdfpages}

\newcommand{\wh}[1]{{\widehat{#1}}} 
\renewcommand{\u}{\mathbf{u}}
\newcommand{\Deltam}{\Delta m} 
\title{Learning stochastic filtering}
\shorttitle{Learning stochastic filtering} 

\author{Rahul O. Ramakrishnan\inst{1} \and  Andrea Auconi\inst{1} \and Benjamin M. Friedrich\inst{1,2}}
\shortauthor{Ramakrishnan \etal}

\institute{                    
  \inst{1} cfaed, Technische Universit\"at Dresden, 01069 Dresden, Germany\\
  \inst{2} Cluster of Excellence ``Physics of Life'', 01307 Dresden, Germany
}

\abstract{
We quantify the performance of approximations to stochastic filtering 
by the Kullback-Leibler divergence to the optimal Bayesian filter. 
Using a two-state Markov process that drives a Brownian measurement process as prototypical test case, 
we compare two stochastic filtering approximations: 
a static low-pass filter as baseline, and machine learning of Voltera expansions using nonlinear Vector Auto Regression (nVAR).
We highlight the crucial role of the chosen performance metric, 
and present two solutions to the specific challenge of predicting a likelihood bounded between $0$ and $1$.
}

\begin{document}

\maketitle
\section{Introduction}

Stochastic filtering is the estimation of the current state of a stochastic process based on a history of noisy measurements. 
The optimal filter uses Bayes formula with the true measurement probabilities to continuously update the likelihood of states. 
If the stochastic dynamics of the underlying process is linear and measurement noise is Gaussian, 
this optimal filter is the celebrated Kalman filter~\cite{kalman1960}. 
For nonlinear problems, numerical approximations are available in the statistics literature \cite{petris2009dynamic}.
These approximations become especially important if the true measurement probabilities are not known.
However, approximations often lack interpretability and it may be difficult to evaluate their performance.

Given a generic stochastic filtering scheme, the uncertainty of its estimates can be decomposed into two contributions: the entropy of the optimal filter, and the Kullback-Leibler (KL) divergence 
\cite{amari2016information,cover1999elements} from the optimal filter. This second term was already suggested as a general measure of model quality in \cite{bialek2020makes};
here, we propose it as a performance measure for stochastic filtering approximations
that track a time-dependent probability distribution.

From an engineering point of view, 
stochastic filtering can be viewed as a problem of system identification: 
find an approximative model of a dynamical system from its time series~\cite{wang2016, abd2020}. 
A priori knowledge about the system allows to make an informed guess, i.e., 
a heuristic model with few parameters that can be tuned to an optimum. 
In the absence of such knowledge, 
one may harness machine learning to learn `black-box' approximation with many parameters ~\cite{steven2016, cheng2017}.
We will demonstrate the applicability of both approaches on a test problem,
and highlight the specific challenges of predicting a probability that is bounded between zero and one, 
as well as possible solutions.

\section{Stochastic filtering}
We start by reviewing the basic features of stochastic filtering.
Let $x_j$ be the unknown state of a stochastic dynamical system at discrete time $j$, 
and $m_j$ a (cumulative) measurement process, whose stochastic increments $\Deltam_j= m_j-m_{j-1}$ 
depend on the state $x_j$ of the hidden process.
The \textit{likelihood} $p(x_j)$ formalizes the knowledge at time $t_j$ about the current state of the hidden process $x_j$ [conditional to all past measurements and possibly an initial prior $p(x_0)$].

The likelihood evolution is decomposed into a \textit{prediction step}, 
$p(x_{j+1})=\int dx_j \, p(x_{j+1}|x_j) p(x_j)$, 
which replicates the stochastic dynamics $p(x_{j+1}|x_j)$ of the hidden process, 
and an \textit{update step}, 
$p(x_{j+1}|\Deltam_{j+1})=p(\Deltam_{j+1}|x_{j+1})p(x_{j+1})/p(\Deltam_{j+1})$, 
which uses Bayes formula to incorporate the latest measurement $\Deltam_{j+1}$.
Here, the measurement probability 
$p(\Deltam_{j+1})=\int dx_{j+1}\, p(\Deltam_{j+1}|x_{j+1}) p(x_{j+1})$ 
acts as normalization factor.
If $p(\Deltam_{j+1}|x_{j+1})$ is the true measurement probability, 
the filter 
$p$
is \textit{optimal}.
The posterior $p(x_{j+1}|\Deltam_{j+1})$ serves as new prior for the next time step ${j+2}$. 
In the limit of a vanishing discretization, 
this procedure yields a time-continuous version of stochastic filtering.
Below, we will assume that the dynamics of the hidden process is stationary, 
and that the influence of the initial prior  vanishes in the long-time limit. 

\section{Model quality}
In many application cases, one has to resort to approximations of optimal stochastic filtering.
The measurement process may not be known, or 
agents may lack the elaborate machinery needed for optimal filtering,
as common for biological cells performing chemical sensing \cite{bialek2012biophysics,kobayashi2010implementation}. 
In these cases, we have a sub-optimal filter, 
whose likelihood estimates $q(x_j)$ will in general differ from those of the optimal filter,
$q(x_j)\neq p(x_j)$.
The performance of this filter $q(x_j)$ can be sub-optimal 
due to reduced dimensionality, intrinsic noise, or systematic errors.

To quantify differences in performance, we examine 
relative entropies. 
The \textit{stochastic entropy} for the optimal filter, $s(x) = - \ln p(x)$, 
is the negative log-likelihood of the actual state $x$. 
Its ensemble average defines the \textit{macroscopic entropy}, 
$S[p] = \langle s(x) \rangle_p = \int \!dx\, p(x) s(x)$, 
which quantifies the current uncertainty of the optimal filter.
The stochastic entropy can also be defined for the sub-optimal estimate, $s^{(q)} (x) = -\ln q(x)$.
Its ensemble average with respect to the true density is the \textit{cross-entropy},
$H[p, q] = \langle s^{(q)}(x) \rangle_p = \int \! dx\, p(x) s^{(q)}(x)$, 
which quantifies the true uncertainty of the sub-optimal filter. 
In general, $H[p, q]$ is different from the macroscopic entropy
$S[q] = \langle s^{(q)}(x) \rangle_q$ of $q(x)$, 
because the sub-optimal filter has an erroneous estimate of its own uncertainty.

Optimal and sub-optimal estimates are related by a fluctuation-theorem,
$\left\langle \exp \left[ s(x) - s^{(q)} (x) \right] \right\rangle_p = 1$,
which follows from the normalization condition for $q(x)$.
Using Jensen's inequality, we obtain the standard upper bound on the entropy by the cross-entropy, 
$S[p] \leq H[p,q]$, 
with equality only if $p=q$. 
This means that the log-likelihood of the actual state evaluated by the optimal filter is on average equal or greater than the log-likelihood of any other scheme, which we would indeed name sub-optimal. Still, the fluctuation theorem allows that in some realizations one can observe apparent violations $p(x) < q(x)$.

The excess uncertainty of the sub-optimal filter is the \textit{Kullback-Leibler divergence} 
$D_{KL}[p\Vert q] = H[p,q]-S[p]$, 
which gives
$D_{KL}[p\Vert q] = \langle \ln(p/q) \rangle_p$. 
The Kullback-Leibler divergence was already proposed to measure model quality in \cite{bialek2020makes}.
Here, we propose to use a time-average $\langle D_{KL}[p\Vert q]\rangle$ 
to quantify the performance of sub-optimal filters.

Below, we will consider a discrete state space for the hidden process $x$ composed of just two states $x{=}\pm 1$.
By slight abuse of notation, we can identify 
$p$
with the scalar $p(x{=}1)$, 
and similarly $q$ with $q(x{=}1)$, and write 
\begin{equation}
D_{KL}[p||q] = p\,\ln\left(\frac{p}{q}\right) \,+(1-p)\,\ln\left(\frac{1-p}{1-q} \right) \quad.
\label{eq:DKL}
\end{equation}
Next, we introduce a minimal bistable model of stochastic filtering, 
and apply eq.~\eqref{eq:DKL} to quantify the performance of sub-optimal approximations.

\section{Bistable model}
\begin{figure*}
  \centering
  \begin{subfigure}[b]{0.42\textwidth}
    \includegraphics[trim={0cm 0.2cm 0 0}, clip, width=\linewidth]{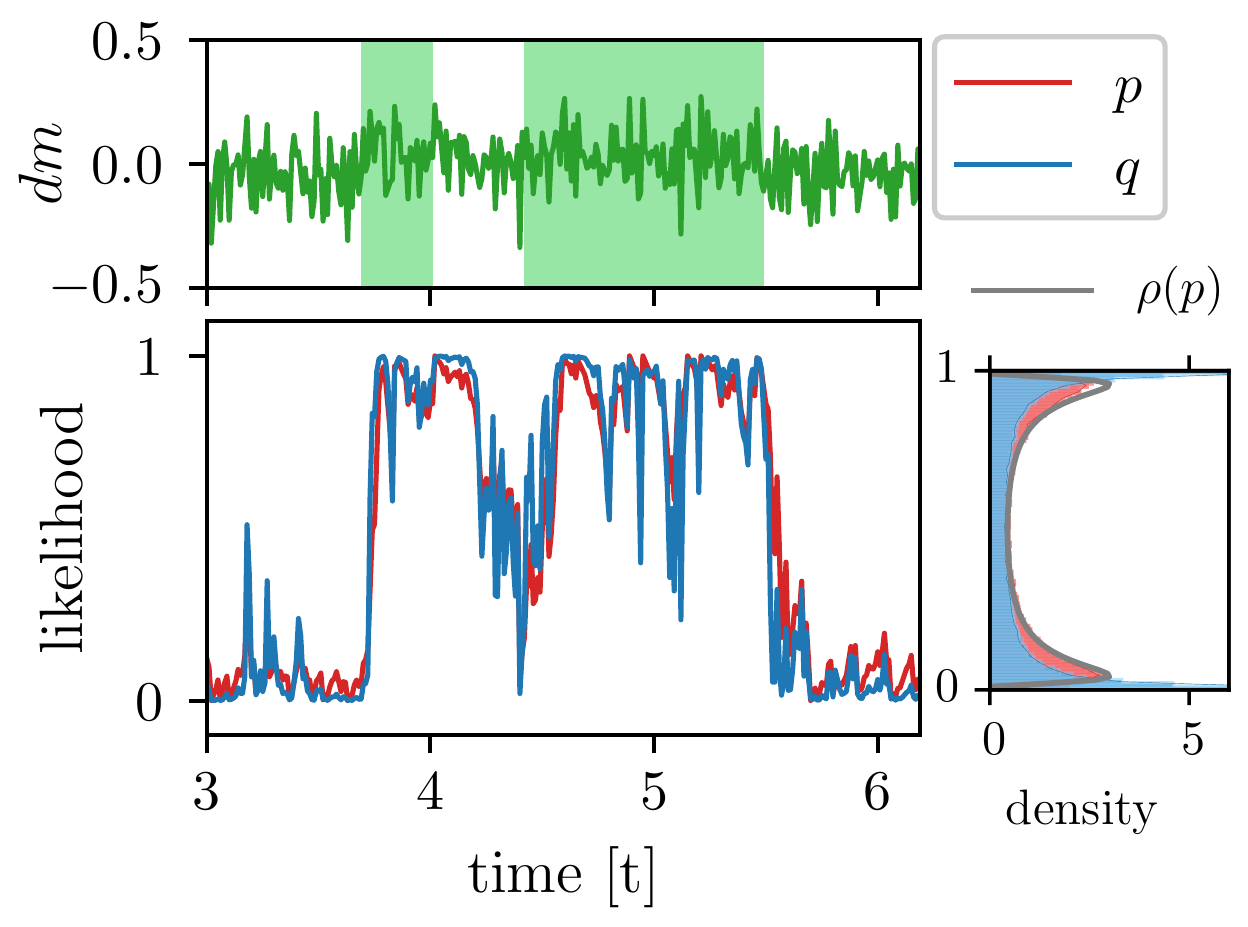}
    \caption{Stochastic filtering of a noisy two-state process}
  \end{subfigure}
  \begin{subfigure}[b]{0.28\textwidth}
    \includegraphics[trim={0.1cm 0 0.4cm 0cm}, clip,width=1.\linewidth]{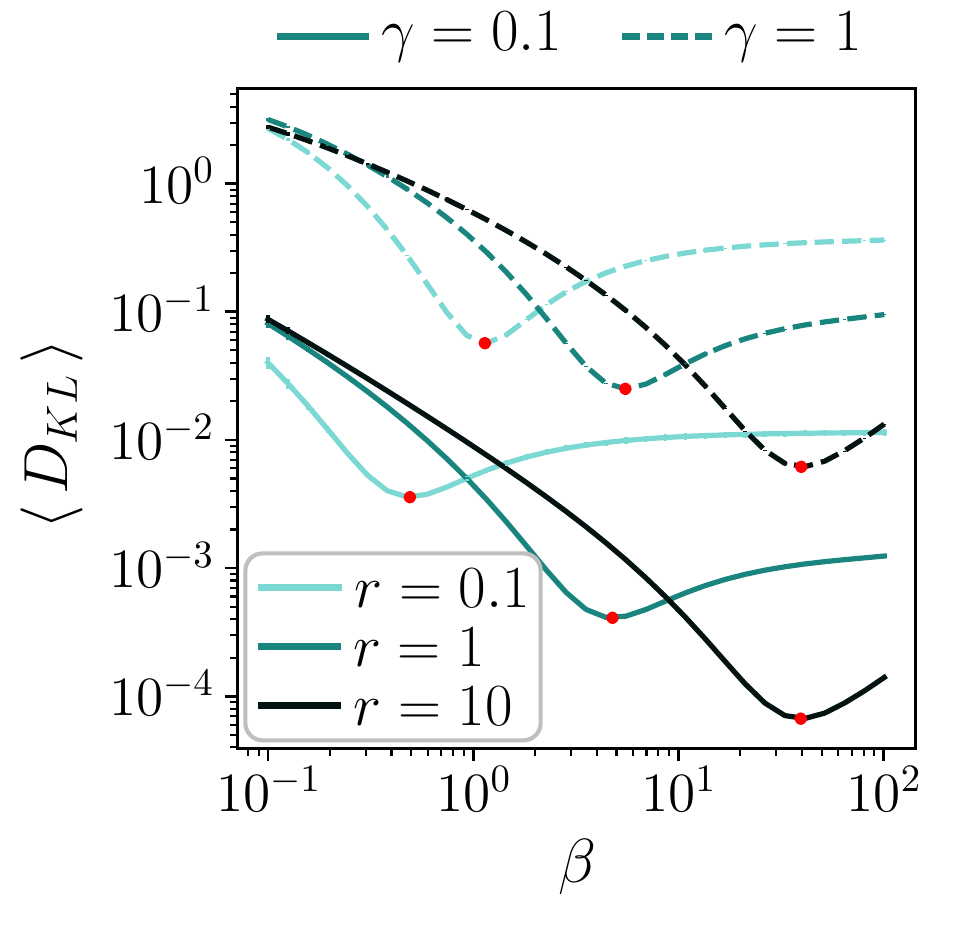}
    \caption{Kullback-Leibler divergence}
  \end{subfigure}
  \begin{subfigure}[b]{0.28\textwidth}
    \includegraphics[trim={0 0 0.3cm 0cm}, clip, width=\linewidth]{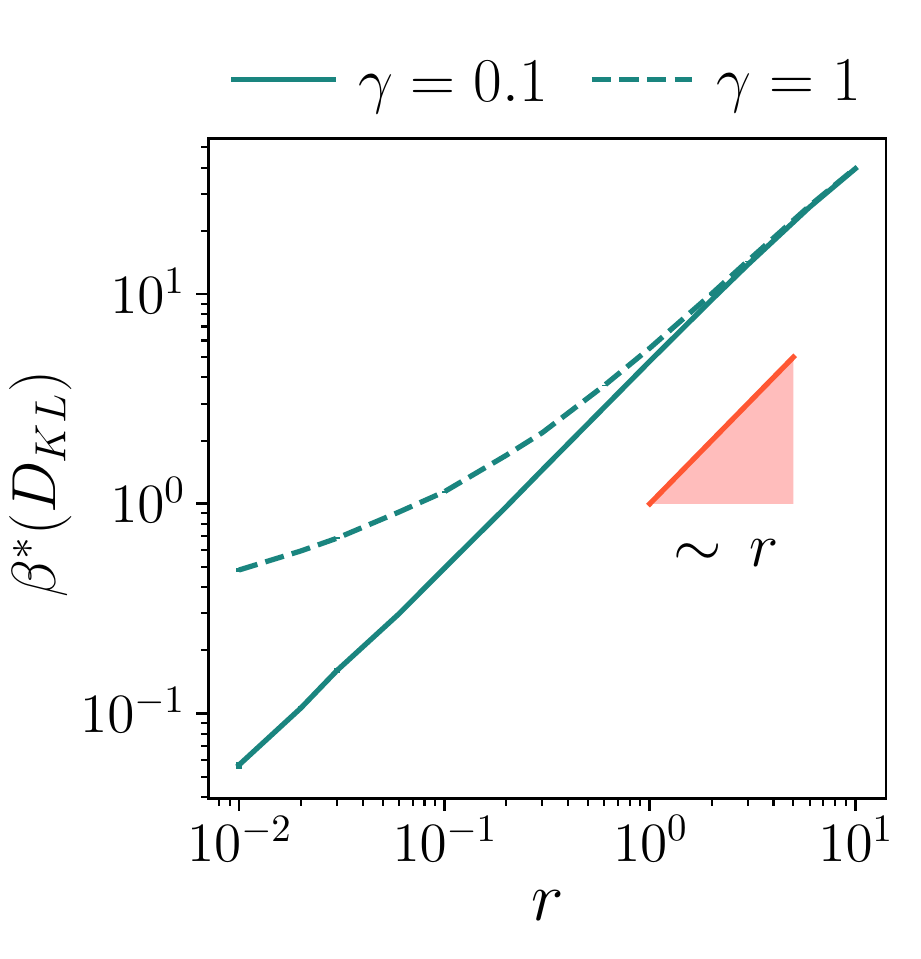}
    \caption{Optimal relaxation time}
  \end{subfigure}
\caption{
(a) 
As prototypical stochastic filtering task, 
we consider a Markov process $x(t)$ jumping between two states
$+1$ (upper panel, green shading) and $-1$ (no shading) that
drives a noisy measurement process $dm$, see eq.~\eqref{measurement model}. 
Bayesian inference yields an optimal estimate of the likelihood $p(t)$ of $x(t){=}{+}1$ (lower panel, red curve), 
see eq.~\eqref{dp}.
As a sub-optimal approximation $q(t)$, we consider a low-pass filter (blue curve), eq.~\eqref{logistic}.
Histogram of these likelihood estimates are shown to the right. 
(b) 
Mean Kullback-Leibler divergence $\langle D_{KL}[p\Vert q]\rangle$ 
between optimal and sub-optimal filter as function of 
the inverse relaxation time $\beta$ of the sub-optimal low-pass filter for different 
noise strengths $\gamma$ and switching rates $r$ of $x(t)$
(where $\beta$, $\gamma$, and $r$ are measured relative to $2D=1$). 
(c)
There exists an optimal inverse time-scale $\beta^\ast$ 
minimizing $\langle D_{KL}[p\Vert q]\rangle$, 
which increases with $r$. 
Results are shown for two values of $\gamma$, 
corresponding to low signal strength ($\gamma=0.1$, solid), and
high signal strength ($\gamma=1$, dashed).
Error bars given by standard error of the mean ($n=5$ realizations) are virtually invisible.
}
\label{lowpass}
\end{figure*}

Consider a continuous-time Markov process $x$ that jumps with symmetric rate $r \geq 0$ 
between two states, $x\in \{-1,+1\}$, see fig.~\ref{lowpass}(a).
Let us assume that we can perform measurements of the hidden state $x_t$ through a continuous measurement process $m_t$ described by the following stochastic differential equation (SDE),
\begin{equation}\label{measurement model}
dm = \gamma \,x\, dt + \sqrt{2D}\,dW,
\end{equation}
where $\gamma\geq0$ is the signal strength
and $\sqrt{2D}\, dW$ denotes Gaussian white noise with noise strength $D$.
The measurement model of eq.~\eqref{measurement model} 
describes, e.g., activation of receptors in a biological cell
exposed to a time-varying, bistable ligand concentration $x$
in the presence of stochastic ligand-receptor interactions subsumed as measurement noise
\cite{kobayashi2010implementation,siggia2013decisions}.

We denote by $p$ the likelihood of state $x=1$.
The optimal Bayesian estimation scheme allows to derive 
evolution equations for quantities of interest, 
such as the time-dependent likelihood $p(x,t)$, 
its invariant density $\rho(p)$, or the expected entropy variation $\langle dS\rangle/dt$.
The calculation methods are standard and can be found, e.g., in 
\cite{kobayashi2010implementation,siggia2013decisions} (likelihood evolution)
\cite{fujisaki1972stochastic,rogers2000diffusions} (invariant density)
\cite{auconi2021gradient} (entropy variation). 

Here, we report analytical results for the specific application problem embodied in eq.~\eqref{dp};
for the convenience of the reader, details of the calculation are provided in the Supplementary Information (SI)
\begin{equation}
\label{dp}
dp = \frac{\gamma}{D}\, p(1-p) \left( dm-\langle dm \rangle_p \right) + r(1-2p)\,dt \quad.
\end{equation}
Here, 
$\langle dm \rangle_p = \gamma (2p-1) dt$, 
and products with the measurement process $dm$ are to be interpreted in the 
It\=o scheme of non-anticipative stochastic calculus.

The invariant density of this likelihood reads
\begin{equation}\label{invariant}
\rho(p)=\frac{K}{p^2 (1-p)^2}\exp\left[-\frac{r D}{\gamma^2}\frac{(2p-1)^2}{p (1-p)}\right] \quad, 
\end{equation}
where $K$ is the normalization constant, see SI Notes. 
This invariant density is bimodal;
implications of this bimodality for cellular signaling were studied in \cite{kobayashi2011connection}.

Similarly, eq.~\eqref{dp} also implies an evolution equation for the entropy $S[p]$, 
and, in particular, for the expected entropy variation with respect to the current likelihood
\begin{equation}
\label{dS}
\frac{\langle dS \rangle}{dt}= r(1-2p)\ln\left(\frac{1-p}{p}\right) - \frac{\gamma^2}{D} p (1-p)\quad. 
\end{equation}
The first term in eq.~\eqref{dS} describes the expected increase of entropy in the prediction step, 
while the second term describes the expected decrease of entropy in the update step.
Both terms depend on $p$ in a nonlinear fashion, and are invariant under the symmetry operation $p\rightarrow 1-p$.
The prefactor $\gamma^2/D$ of the second term can be interpreted as a rate constant of information gain, similar to \cite{mora2019physical,novak2021bayesian}. 
The thermodynamic interpretation of entropy variations in stochastic systems with measurement components has been discussed in 
\cite{sagawa2012nonequilibrium,ito2013information,bartolotta2016bayesian,auconi2019information}.

\section{Approximation I: Low-pass filter}
To construct our first approximation,
consider a low-pass filter of the measurement process, 
$d\xi = dm-\beta\, \xi\,dt$, 
with inverse relaxation time $\beta$.
We want to estimate a likelihood $q(\xi)$ for $x_t{=}{+}1$ based on $\xi_t$.
In the limit of rare switching $r\ll \beta$, 
the probability density $p(\xi|x)$ is given by 
the steady-state density of an Ornstein-Uhlenbeck process: 
a normal distribution with mean $\gamma x/\beta$ and variance $D/\beta$.
By treating the hidden state as static, we introduce a systematic bias.
We can now apply Bayes formula using this steady-state density and the prior $p(x{=}{+}1)=1/2$, 
and obtain the approximated likelihood
\begin{equation}\label{logistic}
q\left(x_t{=}{+}1\big|\xi_t\right)=\frac{1}{1+\exp(-2\gamma\, \xi_t / D)} \quad,
\end{equation}
which describes a logistic curve. 
Applying It\=o's Lemma to eq.~\eqref{logistic}, 
we obtain an evolution equation for the approximated likelihood
\begin{multline}
dq = \frac{2\gamma}{D} 
q(1-q)\left(dm-2\langle dm \rangle_q \right) \\[-3mm]
 -\beta q(1-q)\ln\left( \frac{q}{1-q} \right) dt,
\end{multline}
where $\langle dm \rangle_q =\gamma (2q-1) dt$ 
is the expectation value of the measurement increment according to the approximated likelihood~$q$.

We numerically determined the performance of this approximative stochastic filter 
using the Kullback-Leibler divergence, see fig.~\ref{lowpass}(b).
Higher signal strengths $\gamma$ reduce the performance relative to the optimal filter (higher $D_{KL}$), 
because likelihoods close to zero and one become more frequent, 
which are more difficult to approximate.
Similarly, a lower switching rate $r$ makes the estimation problem easier 
for both the sub-optimal and the optimal filter, 
but $D_{KL}[q\Vert p]$ may nonetheless increase, 
as only the optimal filter 
has the correct functional dependence on the measurement process.
Lastly, the inverse relaxation time $\beta^\ast$ of the low-pass filter 
marks a trade-off between responding \textit{fast} (small $\beta$) or responding \textit{precisely} (high $\beta$), 
resulting in an optimum $\beta^\ast$ that minimizes the Kullback-Leibler divergence.
This optimal $\beta^\ast$ increases with the switching rate $r$ as expected,
with approximately linear scaling for low $\gamma$, see fig.~\ref{lowpass}(c).

\begin{figure*}
  \centering
  \begin{subfigure}[b]{0.65\textwidth}
    \includegraphics[width=\linewidth]{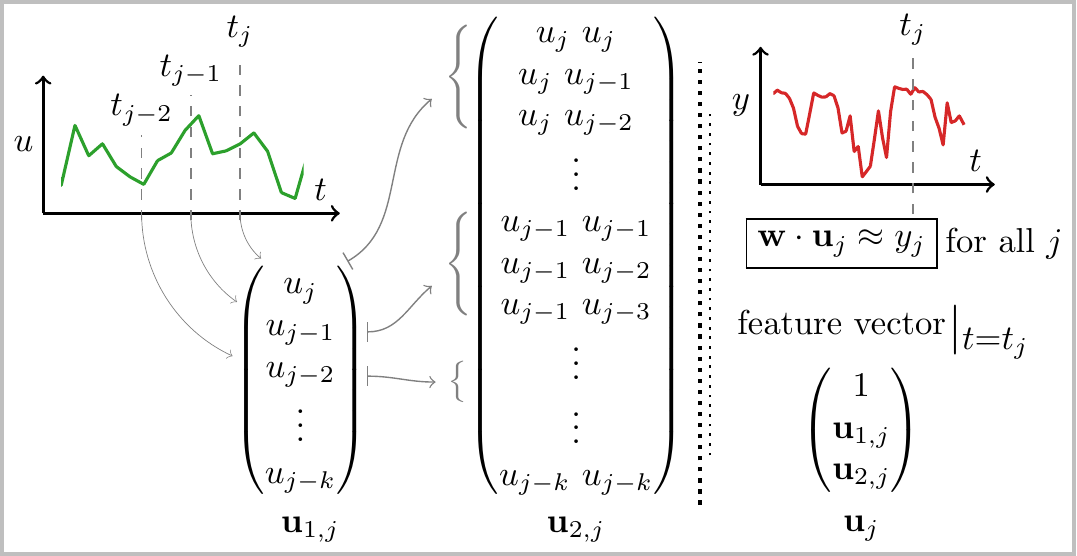}
    \caption{nVAR machine}
  \end{subfigure}
  \begin{subfigure}[b]{0.34\textwidth}
    \includegraphics[trim={0 0 0.2cm 0}, clip, width=\linewidth]{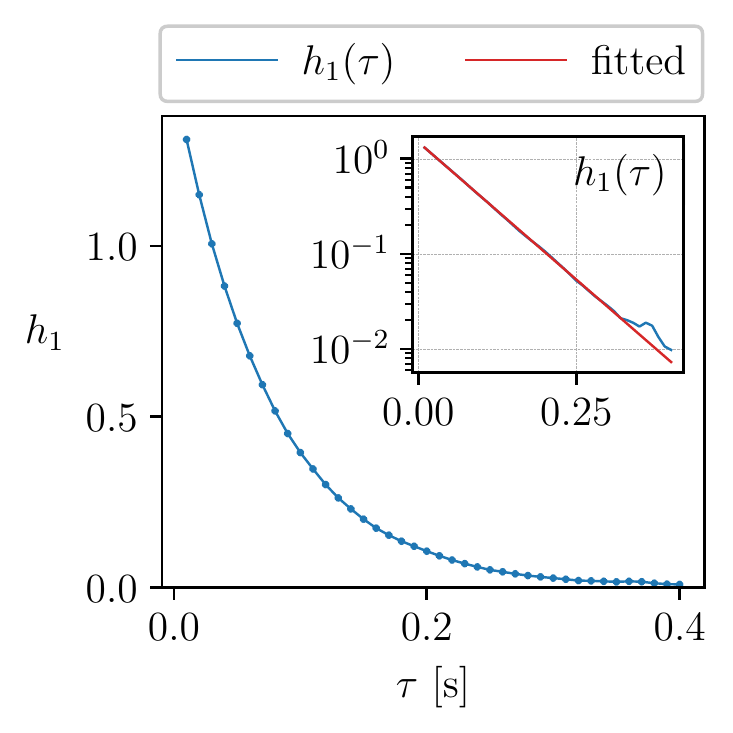}
    \caption{Linear kernel from weights}
  \end{subfigure}
  \caption{
(a) 
nVAR is a method to approximate arbitrary dynamic relationships between a time-dependent input $u(t)$ 
and an output $y(t)$
by learning the kernels of a truncated Volterra expansion, eq.~\eqref{volterra_operator}. 
During training, for each time point $t_j$, feature vectors $\u_{n,j}$ 
are constructed from monomials of order $n$ in the time-delayed inputs $u(t_j)$, $u(t_{j-1})$, ..., $u(t_{j-k})$.
These feature vectors (up to maximal order $n\le N$) are combined into a single feature vector $\u_j$.
All $\u_j$ are then concatenated horizontally into one feature matrix $\mathbf{U}$.
Linear regression with regularization (ridge regression), eq.~\eqref{eq:min}, 
yields an optimal weight vector $\mathbf{w}$ 
such that the vector $\mathbf{y}$ of outputs $y(t_j)$ 
is approximated as $\mathbf{y} \approx \mathbf{w}\cdot\mathbf{U}$, 
i.e., $y(t_j) \approx \mathbf{w}\cdot\mathbf{u}_j$ for all $j$.
(b) 
Application of nVAR to the stochastic filtering problem of fig.~\ref{lowpass}(a) 
with $N=3$, equivalent to learning the kernels $h_1$, $h_2$ and $h_3$ of a Volterra expansion;
$h_1(\tau)$ is shown (blue curve, average of $10$ realizations). 
Approximately, 
$h_1(\tau) \approx A\, e^{-\beta\tau}$
with fit parameters $A\approx 3/2$ and $\beta\approx 40/3$
(red curve in inset).
Parameters: 
$\gamma=3\,\mathrm{s}^{-1}$, 
$r=3\,\mathrm{s}^{-1}$, 
$D=0.5\,\mathrm{s}^{-1}$, 
duration of training data $800\,\mathrm{s}$, 
$\Delta t = 0.01\,\mathrm{s}$, 
delay $k = 40$, $N=3$,
ridge parameter $\alpha = 10^{-3}$.
}
\label{fignvar}
\end{figure*}

\section{Approximation II: machine learning}
As a second approximation of stochastic filtering, 
we consider nonlinear vector auto regression (nVAR)~\cite{gauthier2021next},
a state-of-the-art machine learning approach 
that learns, e.g., Volterra expansions from time series data.
After briefly reviewing the theory behind nVAR, 
we propose how nVAR can be adapted for the problem of stochastic filtering, 
and demonstrate its performance for this task.

Generally, any dynamical system with input $u(t)$ and output $y(t)$ 
can be thought of as a functional relationship 
$y(t) = \mathcal{H}\, [u(t)]$
for some functional $\mathcal{H}$ that operates on time-dependent functions.
For linear and time-invariant systems, 
one can show that this input-output relationship is just a convolution 
\begin{equation}
y(t) = h_0 + \int_{-\infty}^{\infty}  h_1(\tau) \, u(t-\tau) d\tau
\label{h1}
\end{equation} 
with a constant $h_0$ and a suitable kernel $h_1(\tau)$
(sometimes called \textit{linear response function} or susceptibility).
Causality implies $h_1(\tau)= 0$ for $\tau<0$.
For non-linear systems, eq.~(\ref{h1}) generalizes to the \textit{Volterra series expansion}, 
a well known description in system identification \cite{franz2006unifying} 
\begin{equation}\label{volterra_operator}
y(t) = H_0 [u(t)] + H_1 [u(t)] + H_2 [u(t)] + ... + H_n [u(t)] + \ldots \quad.
\end{equation}
Here,
$H_0 [u(t)] = h_0$, and 
$H_n$ is the $n$th-order Volterra operator, which may be written as a ``higher-order convolution''
\begin{multline}
H_n [u(t)] = \int_{-\infty}^{\infty} \cdots \int_{-\infty}^{\infty} h_{n} (\tau_1, \ldots,  \tau_n) \\ 
  \ u(t-\tau_1) \ldots \ u(t-\tau_n) \ d\tau_1, \ldots, d\tau_n \quad.
\label{volterra}
\end{multline}
Theoretically, inputs from a distant past and higher-order kernels influence the output $y(t)$.
For practical purposes, however, we can only use a finite input history and must truncate the Volterra expansion at some order. 
Despite these limitation, learning Volterra kernels using nVAR could approximate even non-polynomial systems surprisingly well~\cite{gauthier2021next}.

Fig.~\ref{fignvar} outlines the nVAR algorithm: 
The training data consists of discrete input and output time series $\{u(t_j)\}$ and $\{y(t_j)\}$.
For each time point $t_j=j\,\Delta t$, 
a \textit{feature vector} $\u_j$ is constructed,
whose components comprise all unique monomials (up to a given order $N$) 
built from the $k+1$ delay terms $u(t_j)$, $u(t_{j-1})$, ..., $u(t_{j-k})$.
The computational cost scales as $\mathcal{O}(k^N)$.
The feature vectors $\u_j$ are then concatenated into a matrix $\mathbf{U}$, which, 
in direct analogy to the Volterra expansion, 
should be linearly related to the \textit{output vector} $\mathbf{y}^{\text{target}}$
(whose components are the $y(t_j)$)
by a yet unknown \textit{weight vector} $\mathbf{w}$ as
$\mathbf{w} \cdot \mathbf{U} \approx \mathbf{y}^{\text{target}}$.
The optimal weights $\mathbf{w}$ are found by the minimization
\begin{equation}
\Vert \mathbf{w} \cdot \mathbf{U} - \mathbf{y}^{\text{target}} \Vert^2 
+ \alpha\, \Vert \mathbf{w} \Vert^2 
\rightarrow \text{min} \quad,
\label{eq:min}
\end{equation}
where $\Vert\cdot\Vert$ stands for the Euclidean norm. 
Eq.~(\ref{eq:min}) includes a penalty for large weights with so-called \textit{ridge parameter} $\alpha$
to reduce over-fitting.
The optimization problem eq.~(\ref{eq:min}) represents a standard task of quadratic programming, 
and can be efficiently solved by reduction to a linear-algebra problem with formal solution~\cite{Roger2020}
\begin{equation}
\mathbf{w} = \mathbf{y}^{\text{target}}\, \mathbf{U}^T (\mathbf{U} \mathbf{U}^T + \alpha\, \mathbb{I})^{-1} \quad.
\label{regression}
\end{equation}
After training, the performance of the nVAR machine can be evaluated using an independent test data set:
Multiplying the learned weight vector with the new feature matrix constructed from the test input yields
a predicted output $\mathbf{y}^{\text{pred}}$,
which is then compared to the true output $\mathbf{y}^{\text{true}}$ of this test data.

We apply nVAR to the specific problem of stochastic filtering.
As example of application, we consider the Brownian measurement process from eq.~\eqref{measurement model}:
the input time series is given by the series of increments 
$u(t_j) = \Deltam_j \equiv m(t_j) - m(t_{j-1})$, 
while the output corresponds to the likelihoods 
$y(t_j)=p(x(t_j){=}{+}1)$.

\begin{figure*}
  \centering
  \begin{subfigure}[b]{0.48\textwidth}
    \includegraphics[trim={0 0.3cm 0 0.7cm}, clip, width=\linewidth]{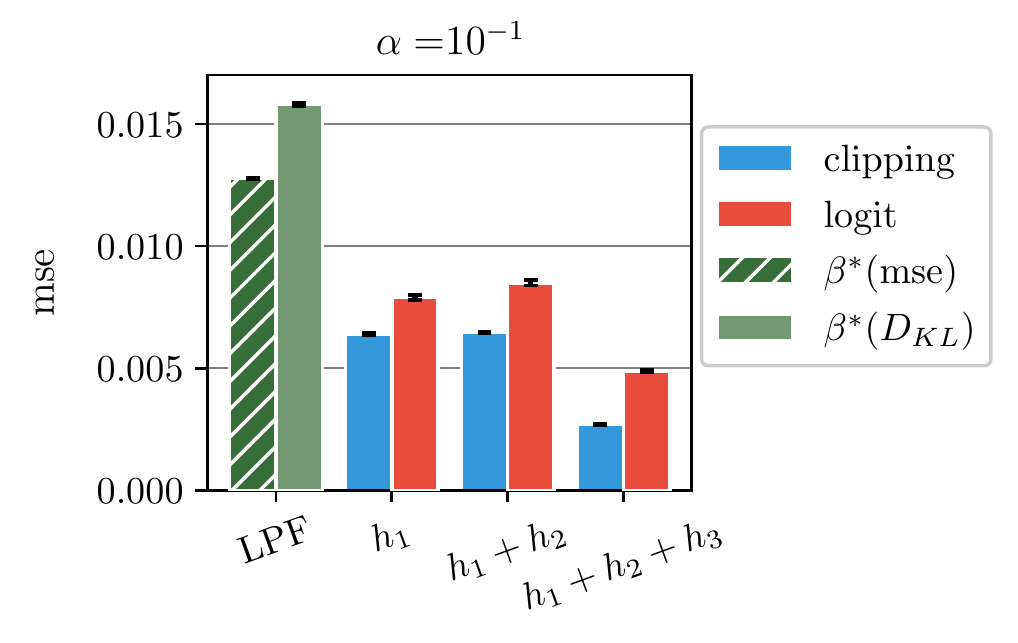}
    \caption{}
  \end{subfigure}
  \begin{subfigure}[b]{0.48\textwidth}
    \includegraphics[trim={0 0.3cm 0 0.7cm}, clip, width=\linewidth]{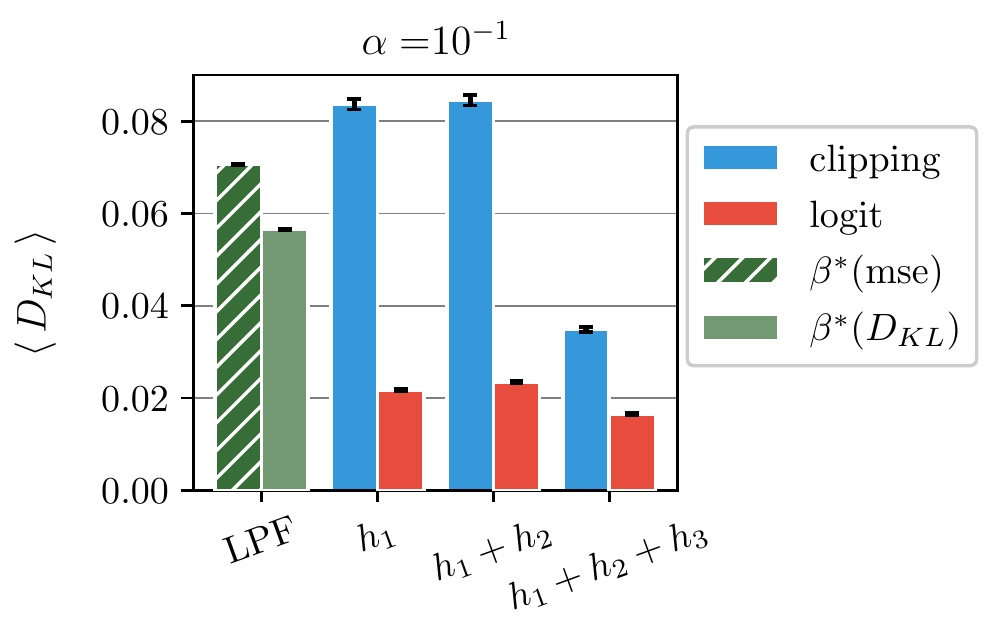}
    \caption{}
  \end{subfigure}
\caption{
Performance of different approximations for stochastic filtering.
(a) 
Performance measured in terms of the mean-squared error (mse) between true and predicted likelihood.
\textbf{LPF:} mse for approximation I using the low-pass filter eq.~\eqref{logistic}. We used the optimal inverse relaxation time $\beta^\ast$
that minimizes either mse (hatched green), or the Kullback-Leibler divergence (as considered in panel (b); pale green).
\textbf{Clipping:} mse for likelihoods predicted using nVAR 
for different truncation orders of the Volterra expansion
($h_1$: $N=1$, $h_1+h_2$: $N=2$, $h_1+h_2+h_3$: $N=3$).
To ensure that likelihoods are bounded within $[0,1]$, 
we chopped predicted likelihoods above $1$ and below $0$ (blue).
\textbf{Logit:}
As an alternative method to ensure that likelihoods lie within $[0,1]$, 
we applied a nonlinear \textit{logit} transformation, eq.~\eqref{logit}, 
to the likelihoods of the training data before training;
the inverse transformation was then applied after prediction, 
but before the mse was computed (red).
(b)
Same as panel (a), but using the mean Kullback-Leibler divergence $\langle D_{KL}\rangle$ as performance measure. 
Here, we use that the likelihood $p=p(x{=}{+}1)$ fully determines the likelihood distribution $p(x)$ for $x{=}\pm 1$
and thus the Kullback-Leibler divergence, eq.~\eqref{eq:DKL}.
Parameters:
$\gamma = 3\,\mathrm{s}^{-1}$, 
$r=3\,\mathrm{s}^{-1}$, 
$D=0.5\,\mathrm{s}^{-1}$,
duration of training time series $800\,\mathrm{s}$, 
duration of test time series $400\,\mathrm{s}$, 
$\Delta t = 0.01\,\mathrm{s}$, 
delay $k = 40$, 
ridge parameter $\alpha = 10^{-1}$; 
reported results represent mean$\pm$s.e.m.
from $10$ realizations.}
\label{barplot}
\end{figure*}

From the learned weights $\mathbf{w}$, 
we can read off the Volterra kernels $h_n(\tau_1,\ldots,\tau_n)$:
The constant term $h_0$ is close to the theoretically expected value $1/2$. 
The weights associated with the linear monomials $\u_1=\{u(t_j),u(t_{j-1}),\ldots \}$ yield, 
after multiplying with $\Delta t$, 
a time-discrete approximation of the first order kernel $h_1(\tau)$,  
see fig.~\ref{fignvar}(b).
The approximated kernel follows an exponential decay 
$h_1(\tau)\approx (\gamma/2) \exp(-\beta\tau)\,\theta(\tau)$,
as expected.
The bistable process considered as example of application 
obeys the symmetry, 
$\mathcal{H}[-x(t)] = 1 - \mathcal{H}[x(t)]$.
This symmetry implies that $h_2(\tau_1,\tau_2)$ 
(and in fact all Volterra kernels of even order) must vanish identically.
In agreement with this symmetry, 
all weights associated with the second-order monomials are nearly zero. 
Moreover, these small weights do not reflect any apparent functional relation, 
suggesting that these weights result from the finite size of the training data.
In line with this interpretation, inclusion of second-order terms does not improve the performance of the stochastic filtering approximation, whereas third-order terms provide a statistically significant improvement, see fig.~\ref{barplot}.

A peculiarity of stochastic filtering is that likelihoods must always lie in the interval $[0,1]$.
\textit{Per se}, the nVAR algorithm  does not respect this property.
To solve this issue, we suggest two different solutions. 
The first, simple solution is to chop the predicted output above $1$ and below $0$. 
This makes computed mean-squared errors more meaningful, 
and is even a prerequisite to compute a Kullback-Leibler divergence.

As a second solution, 
we applied a nonlinear \textit{logit}-transformation to the likelihoods $p$
\begin{equation}
\phi = \text{logit} (p) =  \ln\left(\frac{p}{1-p} \right)\quad,
\label{logit}
\end{equation}
which maps $0<p<1$ bijectively to $-\infty<\phi<\infty$. 
We then use nVAR to predict $\phi(t_j)$ with input $\Deltam_j$;
the predicted output $\wh{\phi}$ is then transformed back to the corresponding likelihood $\wh{p}$ 
using the inverse transformation
\begin{equation}
\wh{p} = \text{logit}^{-1}( \wh{\phi}) = [\, 1+ \exp(-\wh{\phi})\, ]^{-1}\quad.
\label{logitinv}
\end{equation}
It is worth to note the formal similarity between eq.~\eqref{logitinv} 
and the simplified approximation eq.~\eqref{logistic}.
On a practical note, to avoid large values of $\phi$, which may cause problems during learning,
we clipped inputs $p$ to the interval $[\varepsilon, 1-\varepsilon]$ for a small number $\varepsilon = 10^{-8}$
before computing the transformed input $\phi$.

Fig.~\ref{barplot} compares these two approaches to deal with the fact that likelihoods are bounded,
using two different performance measures, 
the conventional mean-squared error (mse), and the mean Kullback-Leibler divergence.
Generally, the first approach (clipping) performs slightly better than the second approach (logit)
in terms of the mean-squared error, which is expected because the first approach directly minimizes this measure.
In contrast, the \textit{logit}-approach is superior when the mean Kullback-Leiber divergence should be minimized.
This observation can be rationalized as follows:
the nonlinear \textit{logit}-transformation eq.~(\ref{logit}) ``stretches'' likelihoods close to $0$ or $1$ out;
as the Kullback-Leibler divergence is rather sensitive to these likelihoods,
the \textit{logit}-approach generally performs better in terms of the Kullback-Leibler divergence.
As discussed above, including second-order terms does not improve performance for any of the two approaches 
as expected by symmetry, whereas third-order terms improve performance.
This improvement is stronger for the clipping-approach compared to the \textit{logit}-approach, 
presumably, because the nonlinear \textit{logit}-transformation partially accounts already for the 
nonlinear dependence of $p$ on the input $dm$.
These observations were confirmed for various parameter choices (not shown).
In the limit of strong noise (small $\gamma$), likelihoods $p$ will deviate only little from $1/2$, 
rendering the dependence of $p$ on $dm$ approximately linear, 
which reduces the benefit of including third-order terms.
Remarkably, the clipping-approach using only linear terms ($h_1$) 
outperforms the simple low-pass filter even if an optimal 
inverse relaxation time is chosen for the later.
This highlights the short-comings of the indirect approach of approximation I, 
which first computes the auxilliary variable $\xi_t$, which is then converted into an estimate for the likelihood using the steady-state probability distribution, 
eq.~\eqref{logistic}, 
instead of learning likelihoods directly as in approximation II.

For machine learning tasks, it is pivotal to choose hyper-parameters judiciously. 
The number $k$ of delay steps should satisfy $k > (\beta^\ast\,dt)^{-1}$ to cover the time-window, 
where the expected Volterra kernels $h_1(\tau)\sim\exp(-\beta^\ast \tau)$ are large.
Fig.~\ref{lowpass}(c) suggests $\beta^\ast\sim r$, where $r$ is the switching rate of the two-state Markov process.

Next, under-fitting or over-fitting will occur
when the number of weights and thus fit parameters is either much smaller or much bigger 
than the size of the training data, respectively.
Intriguingly, performance is worst when the number of weights exactly matches 
the size of the training data \cite{loog2020}.
If the number of weights increases further and exceeds the number of training data points, 
learned weights become partially randomized, which reduces the adverse effect of over-fitting~\cite{Belkin2019}, 
see also SI Notes.
For our application example, performance was best when the number of weights equaled 
about $20\%$ of the number of data points of the training data. 

The common problem of over-fitting and resultant abnormal weights caused by noise in finite-size input data
can be effectively mitigated by means of regularization, i.e., 
introducing a regularization term with ridge parameter $\alpha$ in the optimization problem eq.~\eqref{regression}. 
Prediction performance for various choices of $\alpha$ are shown in the SI Notes.
Interestingly, the \textit{logit}-approach of applying a nonlinear transformation to the output before learning
is rather sensitive to the choice of $\alpha$ if third-order terms are included,
whereas the performance of the simpler clipping-approach is robust and virtually independent of the choice of $\alpha$.

\section{Conclusion} 
We addressed the general problem of stochastic filtering 
to infer a time-dependent hidden state from noisy measurements.
Because the optimal filter based on Bayesian inference can be analytically calculated only for simple examples, 
efficient approximations are needed.
Here, we propose nonlinear vector auto-regression (nVAR)~\cite{gauthier2021next}
as an efficient method to learn the dynamic relationship between 
the likelihood of hidden states and noisy, time-dependent input.
Specifically, nVAR allows to learn a Volterra expansion relating input and output, 
in our case, noisy measurements as input and likelihoods of hidden states as output.
As a baseline, we additionally consider a simple low-pass filter.
We quantify the performance of these approximations of stochastic filtering 
in terms of the Kullback-Leibler divergence from the optimal filter.

Using higher-order terms in the Volterra expansion or longer delays in nVAR, increases the accuracy of the predictor, 
but also increases computational cost. 
We show that nVAR with reasonable delays and only linear terms already performs better than the simple low-pass filter. 
Including third-order terms slightly improves performance further, 
whereas second-order terms are dispensable because of the symmetry of the problem.

A key issue is that likelihoods must always be bounded between zero and one --
a property not automatically respected by common approximation schemes. 
We propose two possible solutions to this specific challenge of stochastic filtering:
(i) clipping of predicted likelihoods outside the admissible range, and
(ii) applying a nonlinear \textit{logit}-transformation before training, and back-transformation of predicted outputs.
The second approach displays higher fidelity when it comes to predicting likelihoods close to zero or one, 
and, concomitantly, shows a lower mean Kullback-Leibler divergence than the first approach. 
At the same time, the mean-squared error resulting from the second approach is still acceptable.
As a drawback, the second approach is sensitive to proper regularization, and 
requires judicious choice of a ridge parameter.
One could think that yet a third solution may be to learn a non-linear model, 
where the output is given, e.g., in terms of a sigmoidal function applied to a weighted sum of the input features.
However, learning such a non-linear model would be considerably harder than learning a linear model like nVAR,
and convergence to a global optimum cannot be guaranteed
(and in fact is unlikely for the larger number of fit parameters used here).

In conclusion, the \textit{logit}-transformation seems a viable approach 
that minimizes the mean Kullback-Leibler divergence, 
while the mean-squared error remains acceptable.
As a caveat, the \textit{logit}-transformation is more susceptible to over-fitting 
and requires suitable regularization,
while the simpler clipping approach is more robust.

While we restricted ourselves in nVAR to learning nonlinear terms up to third order, 
the method could be applied up to arbitrary order or using longer delays, 
with the availability of sufficient training data and computational resources being the only bottle-necks.
Computational complexity could be reduced by careful design of the feature vector, 
e.g., using coarser time-sampling for inputs with longer delays, 
or problem-specific basis functions constructed from the inputs~\cite{steven2016}.
Lasso regression allows to shrink less important weights to zero~\cite{tibsh1996}. 
As an alternative to nVAR, multilayer neural networks have been proposed to learn Volterra expansions, 
with a direct relation between the internal weights of the network 
and Volterra kernels~\cite{wray1994, parker1992, cheng2017}.
We have shown how learning Volterra expansions can be adapted to find stochastic filtering approximations, 
enabling future applications for on-line decision making.

\acknowledgments
We thank all members of the Biological Algorithms group as well as Marc Timme for stimulating discussions.
ROR was supported by the Ministry of Science and Art of the Federal State of Saxony, Germany
through grant 100400118  to Marc Timme and BMF,
financed with tax funds on the basis of the budget adopted by the Saxon State Parliament
(Forschungsprojektf\"orderung Titelgruppe 70 des S\"achsischen Staatsministerium f\"ur Wissenschaft und Kunst).
AA was supported by the Deutsche Forschungsgemeinschaft (DFG, German Research Foundation) 
through grant FR3429/3-1 to BMF.
BMF is supported by the DFG through FR3429/4-1 (Heisenberg grant), 
and under Germany's Excellence Strategy - EXC-2068 - 390729961.
ROR, AA, and BMF acknowledge support through the \textit{Center for Advancing Electronics Dresden} (cfaed).

\bibliographystyle{unsrt}
\bibliography{ms.bib}
\end{document}


\renewcommand{\theequation}{S\arabic{equation}}    
\setcounter{equation}{0}  
\renewcommand{\thefigure}{S\arabic{figure}}    
\setcounter{figure}{0}  
\renewcommand{\thetable}{S\arabic{table}}    
\setcounter{table}{0}  

\newcommand{\red}[1]{{\color{black} #1}}

\newcommand{\wh}[1]{{\widehat{#1}}} 
\renewcommand{\u}{\mathbf{u}}
\newcommand{\Deltam}{\Delta m}

\title{Supplementary Material \\ \Large ``Learning stochastic filtering''\\
\normalsize Rahul O. Ramakrishnan, Andrea Auconi, Benjamin M. Friedrich}
\date{}
\maketitle

\section{Likelihood evolution}
Here, we derive eq.~(3) of the main text. Analogous derivations for minimal models of stochastic filtering can be found, e.g., in \cite{siggia2013decisions,kobayashi2010implementation,mora2019physical}.

Here the hidden state is a Markov process that jumps with symmetric rate $r$ between the two states $x\in \{-1,+1\}$.
The measurement process is $dm=\gamma\,x\,dt + dW$, where $dW$ is a standard Brownian motion.
Consider a regular time discretization of the process with time step $\Delta t$. 
The prior likelihood is $p(x_j)\equiv p(x_j|m_{-\infty:j})$. 
Note that we consider all probabilities here as implicitly conditional on the measurement process trajectory up to time $j$.

The predicted likelihood is simply
\begin{equation}
\label{eq:predict}
	p(x_{j+1}) = p(x_j) + r \left(1-2p(x_j)\right) \Delta t +\mathcal{O}(\Delta t^2)\quad.
\end{equation}
The discretized measurement process increment $\Deltam_{j+1} \equiv m_{j+1}-m_j$ is written
\begin{equation}
	\Deltam_{j+1}  = \gamma x_{j+1} \Delta t +\Delta W_{j+1} +\mathcal{O}(\Delta t^2)\quad,
\end{equation}
where $\Delta W_{j+1}=W_{j+1}-W_j$ is the Brownian increment. The correction $\mathcal{O}(\Delta t^2)$ is true in statistical sense. Indeed, the probability of a hidden process jump within the time interval $(t_j,t_{j+1}]$ is $r\Delta t +\mathcal{O}(\Delta t^2)$, and the upper bound on the correction is $2\gamma\Delta t$ corresponding to a jump close to $t_j$. Then the product of the correction times the probability is upper bounded by a term of order $\mathcal{O}(\Delta t^2)$, and is negligible in the limit $\Delta t\rightarrow 0$ with respect to the $\Delta t$ and $\Delta W$ terms which results in a SDE. 
(Alternatively, we could just define the measurement to be dependent on only $x_{j+1}$ and have no corrections.)

The conditional measurement distribution in discrete time is the Gaussian 
$\mathcal{N}(\gamma x \Delta t,\red{2D}\Delta t)$, which we write as
\begin{equation}
\label{eq:measurement}
	p\left(\Deltam_{j+1} \big| x_{j+1}=\pm 1\right) = K(\Deltam_{j+1}) \exp(\pm \gamma \Deltam_{j+1} \red{/(2D)} )\quad,
\end{equation}
where 
$K(\Deltam_{j+1})\equiv (\red{4}\pi\,\red{D} \Delta t)^{-\frac{1}{2}} \exp\left( 
		-\frac{{\Deltam_{j+1}}^2 +\gamma^2 \Delta t^2}{\red{4D}\,\Delta t}
	\right)$. 
The unconditional measurement probability is
\begin{align}
\label{eq:unconditional}
	p(\Deltam_{j+1}) &= \sum_{x_{j+1}=\pm 1}p(x_{j+1})\, p\left(\Deltam_{j+1} \big| x_{j+1}\right) \nonumber \\
	&= K(\Deltam_{j+1}) \bigg[ p(x_{j+1}{=}{-}1) \exp\left(- \frac{\gamma}{\red{2D}} \Deltam_{j+1}\right) 
                             + p(x_{j+1}{=}{+}1) \exp\left(+ \frac{\gamma}{\red{2D}} \Deltam_{j+1}\right) \bigg]\quad.
\end{align}
Then we apply Bayes formula to update the predicted likelihood $p(x_{j+1})$ into the posterior 
$p(x_{j+1}|\Deltam_{j+1})\equiv p(x_{j+1}|m_{-\infty:j+1})$,
and find using eqs.~\eqref{eq:predict}, \eqref{eq:measurement}, and \eqref{eq:unconditional}
\begin{multline}
	p(x_{j+1}{=}{+}1|\Deltam_{j+1}) = 
	p(x_{j+1}{=}{+}1)\,\frac{p\left(\Deltam_{j+1} \big| x_{j+1}{=}{+}1\right)}{p\left(\Deltam_{j+1}\right)}\\
		=p + r(1-2p)\Delta t
		+\frac{\gamma}{\red{D}} p\left(1-p\right)\left[
			\Deltam_{j+1}
          + \frac{\gamma}{\red{2D}} \left(1-2p\right){\Deltam_{j+1}}^2 \right]
		+\mathcal{O}((\Delta t)^\frac{3}{2},{\Deltam_{j+1}}^\frac{3}{2}),
\end{multline}
where we denoted $p\equiv p(x_j{=}{+}1)$ and expanded expressions. Now we use the  formal expression $dWdW=dt$, of Ito stochastic calculus which means that ${\Deltam_{j+1}}^2$ converges to $\red{2D}\,dt$ under the integral sign~\cite{shreve2004stochastic,auconi2021gradient}. Then in the limit $\Delta t\rightarrow 0$ and in the sense of an Ito integral, we obtain the likelihood evolution equation
\begin{equation}
dp = \frac{\gamma}{\red{D}} p(1-p)(dm-\langle dm \rangle) +r(1-2p)dt\quad,
\end{equation}
where $\langle dm \rangle = \gamma(2p-1)dt$.

\section{Steady-state density}
Here, we review the derivation of the steady-state invariant density, which is based on the innovation process technique  \cite{fujisaki1972stochastic,rogers2000diffusions,kobayashi2011connection}. Let us define the \textit{innovation} process as
\begin{equation}
	d\widetilde{W}\equiv dm -\langle dm \rangle = dm +\gamma (1-2p) dt\quad,
\end{equation}
where the average is meant with respect to the current likelihood $p$.
The innovation process increments have zero expectation, $\langle d\widetilde{W} \rangle =0$, and a linear quadratic variation, $d\widetilde{W}d\widetilde{W}=\red{2D}\, dt$. Then from L\'evy theorem \cite{shreve2004stochastic} it follows that the innovation process $\widetilde{W}$ is a Brownian motion. In light of this, we can rewrite the likelihood evolution, eq.~(3) of main text, as
\begin{equation}
dp = \frac{\gamma}{\red{D}}\, p(1-p)\,d\widetilde{W} +r(1-2p)\,dt\quad.
\end{equation}
Let us now introduce the Wald ratio as $f \equiv \ln\left(\frac{p}{1-p}\right)$, and write its evolution equation with Ito's Lemma
\red{(using \red{$df/dp = [p(1-p)]^{-1}$} and $p=[1+e^{-f}]^{-1}$)}
\begin{equation}
  df = \frac{\gamma}{\red{D}}\, d\widetilde{W} + r \left( e^{-f}+e^f \right) -\frac{\gamma^2}{\red{D}} \frac{1-e^f}{1+e^f}\, dt \quad.
\end{equation}
We see that the transformation has led to a non-multiplicative noise term. The corresponding Fokker-Planck equation reads
\begin{align}
  \partial_t P(f,t) &= 
    \frac{\gamma^2}{\red{D}}\, \partial_f^2 P(f,t) 
    -\partial_f \left[ 
		\left( r \left( e^{-f}+e^f \right) - \frac{\gamma^2}{\red{D}} \frac{1-e^f}{1+e^f} \right)
	P(f,t) \right] \nonumber \\ 
 & \equiv -\partial_f \Phi(f,t) \quad,
\end{align}
where in the last line we wrote the continuity equation in terms of the probability current $\Phi(f,t)$. The steady-state equation $\partial_t P^*(f) =0$ in 1D without jumps implies that the probability current is zero \cite{risken1996fokker}, $\Phi^*(f)=0$. Solving this equation, we get the steady-state distribution of $f$, and then we transform it back to obtain the invariant density for the likelihoods $\rho(p)=\Phi^*(f(p)) \big|\frac{df}{dp} \big|$, which is eq.~(4) in the main text,
\begin{equation}
\rho(p) = \frac{K}{p^2 (1-p)^2}\exp\left[-\frac{r\red{D}\,(2p-1)^2}{\gamma^2\, p (1-p)}\right]\quad,
\end{equation}
where $K$ is the normalization constant ensuring $\int_0^1 dp\,\rho(p) = 1$.

\pagebreak
\section{Volterra kernels from the learned weights}
\begin{figure}[H]
\includegraphics[scale=1]{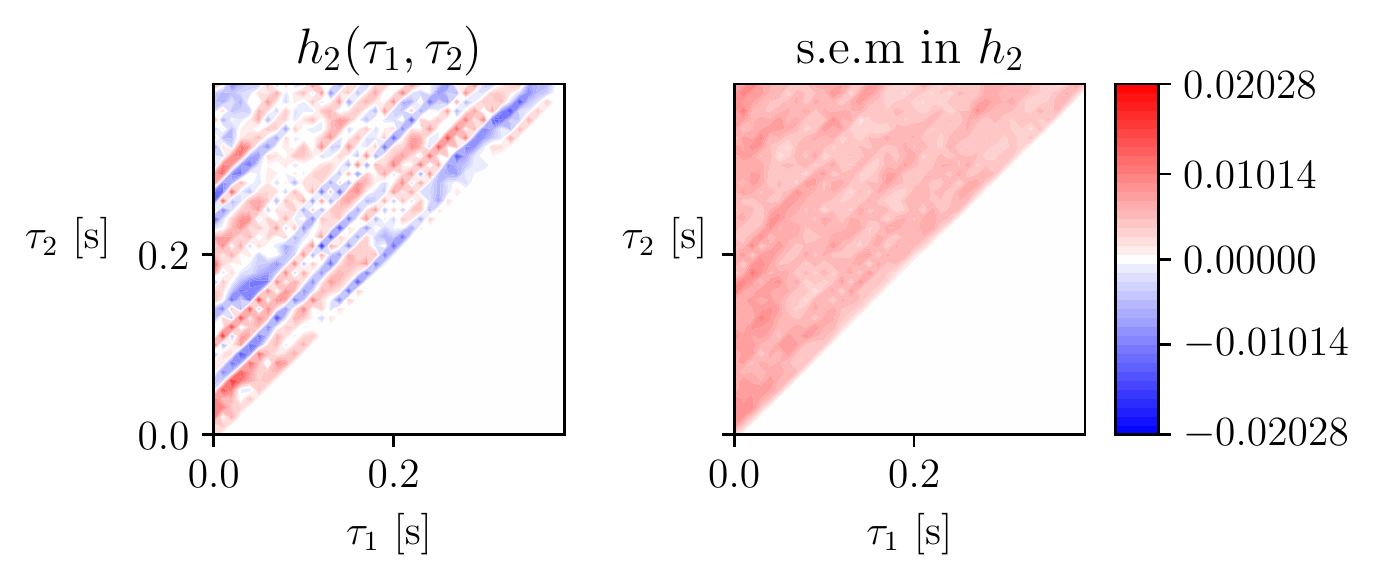}
\caption{Second order Volterra kernel: mean (left), and standard error of the mean (right) from 10 realizations. Note that the nVAR algorithm avoids repetition of the higher order monomials in the feature vector. Hence half of the entries in the $\tau_1 - \tau_2$ plane are empty. 
Parameters: 
$\gamma=3\,\mathrm{s}^{-1}$, 
$r=3\,\mathrm{s}^{-1}$, 
$D=0.5\,\mathrm{s}^{-1}$, 
duration of training data $800\,\mathrm{s}$, 
$\Delta t = 0.01\,\mathrm{s}$, 
delay $k = 40$, $N=3$,
ridge parameter $\alpha = 10^{-3}$.
}
\label{fig_k40h2}
\end{figure}

Here, we visualize the Volterra kernels $h_2$ and $h_3$, obtained from the learned weights. 
The feature vector of the nVAR machine is designed in a peculiar manner as outlined in fig.~2(a) of the main text.
The weights associated with the linear monomials $\mathbf{u}_1$ characterize 
(a time-discrete approximation when multiplying with $\Delta t$) the first order kernel $h_1(\tau)$ as shown in fig.~2(b) of the main text. 
Similarly, one can identify the weights associated with the second order monomials $\mathbf{u}_2$ 
as the second order kernel $h_2(\tau_1, \tau_2)$, see fig.~\ref{fig_k40h2}. 
This kernel is almost zero and does not represent any apparent functional relation. 
Moreover, including $h_2$ in nVAR does not improve the performance of the resultant stochastic filtering approximation. 
Indeed, symmetry implies $h_2 = 0$ in our case.

\begin{figure}[H]
\centering
  \begin{subfigure}[b]{0.45\textwidth}
    \includegraphics[scale=1]{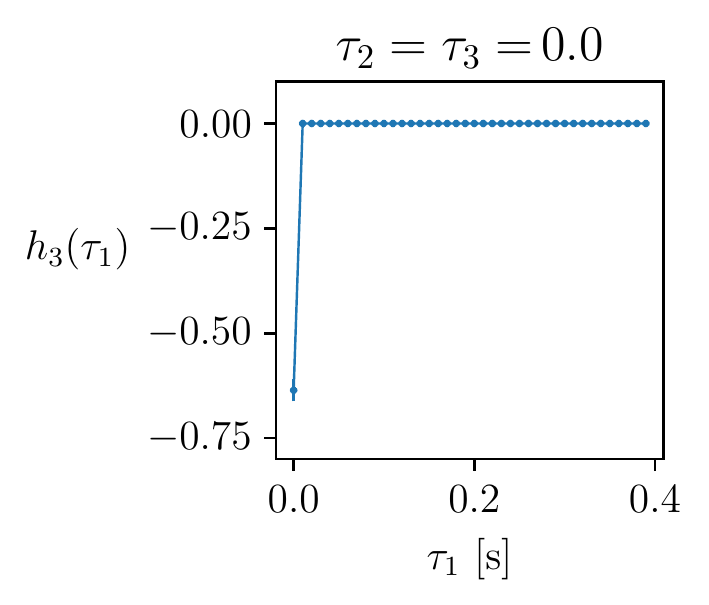}
  \end{subfigure}
  \begin{subfigure}[b]{0.45\textwidth}
    \includegraphics[scale=1]{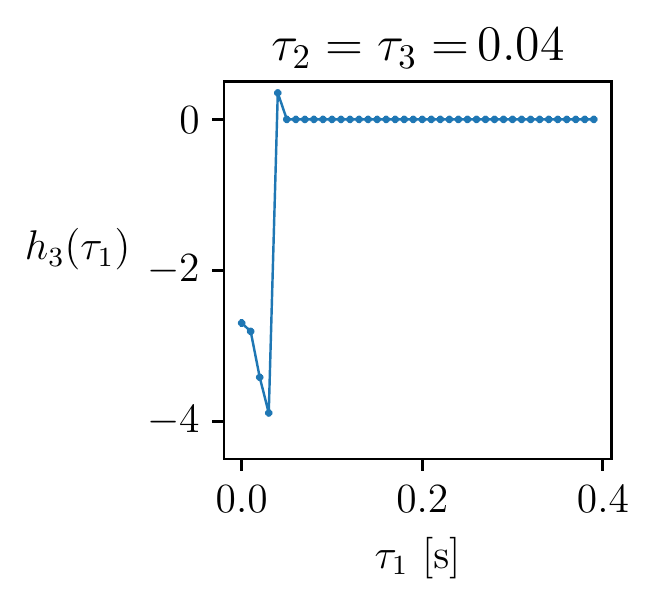}
  \end{subfigure}
  \begin{subfigure}[b]{0.45\textwidth}
    \includegraphics[scale=1]{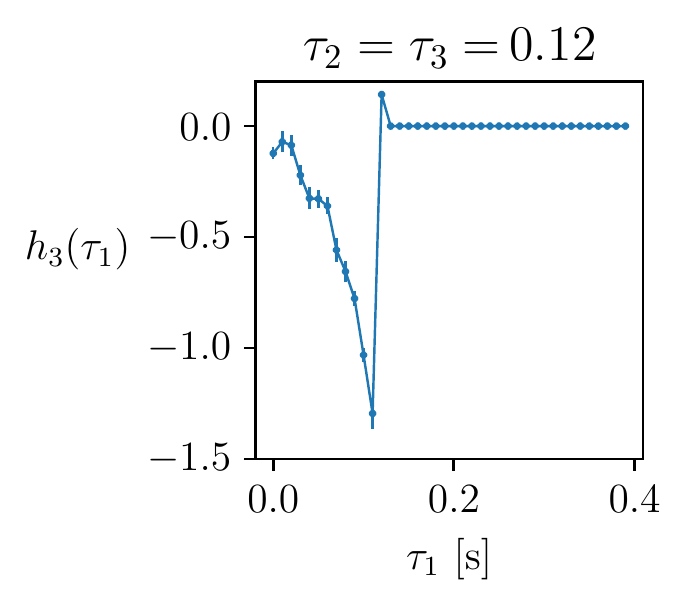}
  \end{subfigure}
  \begin{subfigure}[b]{0.45\textwidth}
    \includegraphics[scale=1]{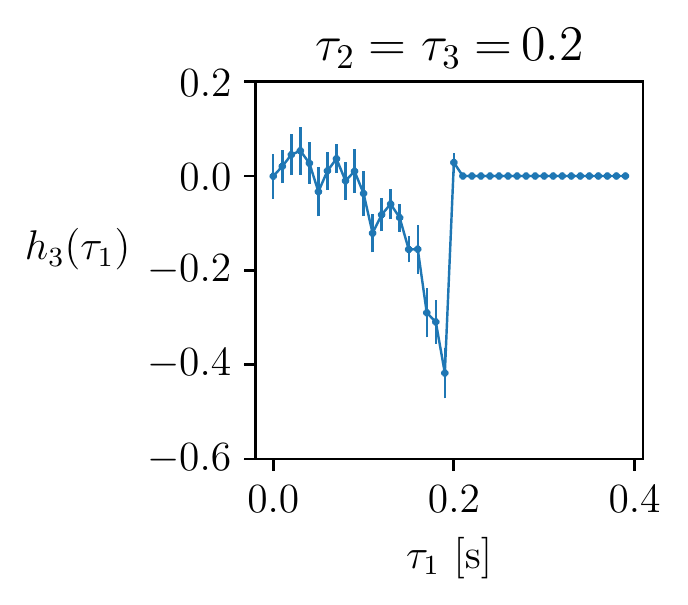}
  \end{subfigure}
\caption{Third order Volterra kernel: visualized as a function of $\tau_1$ for $ \tau_2 = \tau_3 = \text{constant}$. Reported results represent mean$\pm$s.e.m.
from 10 realizations.  Parameters: same as in fig.~\ref{fig_k40h2}}
\label{fig_k40h3}
\end{figure}

Finally, we identify the weights associated with third order monomials $\mathbf{u}_3$ 
as the third order kernel $h_3(\tau_1, \tau_2, \tau_3)$, see fig.~\ref{fig_k40h3}. 
%

\pagebreak

\section{Performance dependence on \textit{ridge parameter}}
\begin{figure}[H]
  \centering
  \begin{subfigure}[t]{0.49\textwidth}
    \includegraphics[width=\linewidth]{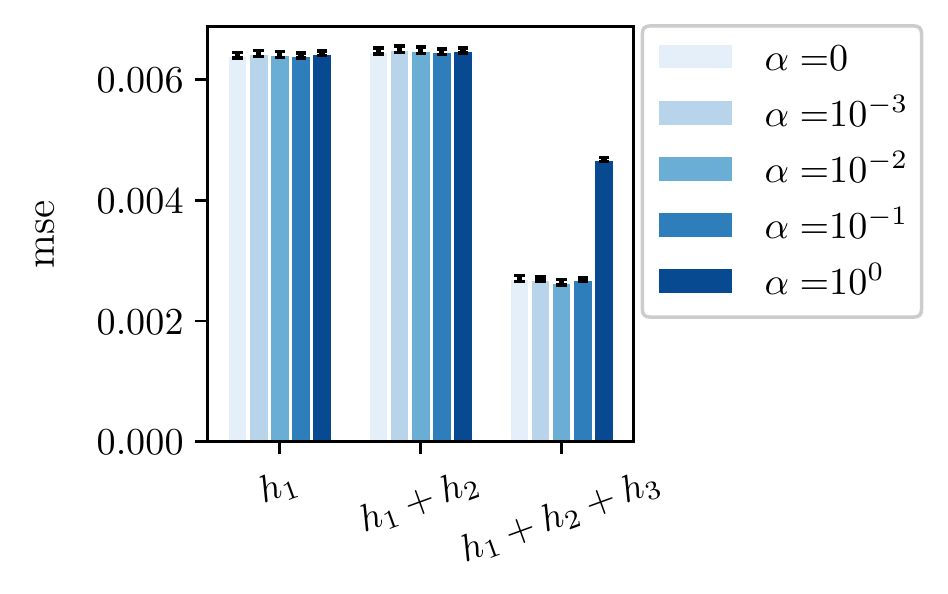}
    \caption{}
  \end{subfigure}
  \begin{subfigure}[t]{0.49\textwidth}
    \includegraphics[width=\linewidth]{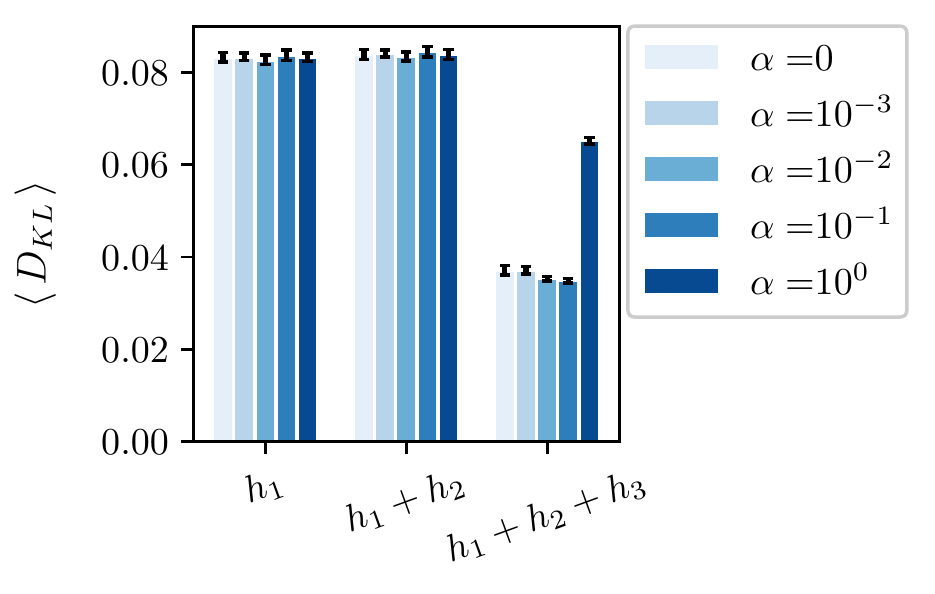}
    \caption{}
  \end{subfigure}
  \caption{\textbf{clipping approach:} performance measures in terms of (a) mean squared error  and (b) mean Kullback-Leibler divergence. Order of the monomials used in the nVAR algorithm is denoted as follows --- ($h_1$: $N=1$, $h_1+h_2$: $N=2$, $h_1+h_2+h_3$: $N=3$). 
Parameters: 
$\gamma=3\,\mathrm{s}^{-1}$, 
$r=3\,\mathrm{s}^{-1}$, 
$D=0.5\,\mathrm{s}^{-1}$, 
duration of training time series $800\,\mathrm{s}$, 
duration of test time series $400\,\mathrm{s}$, 
$\Delta t = 0.01\,\mathrm{s}$, 
delay $k = 40$; 
reported results represent mean$\pm$s.e.m.
from $10$ realizations.}
\label{fig_alpha_clipping}
\end{figure}

In ridge regression, eq.~(12) of the main text, the \textit{ridge parameter} $\alpha$ has to be set judiciously to control the size of optimal weights, as a measure to avoid over-fitting. In this section we show how \textit{ridge parameter} affects the performance in our specific example. We considered two  approaches in our study. Performance measures for the clipping approach is shown in fig.~\ref{fig_alpha_clipping}. The performance virtually did not change over a wide range of $\alpha$.

\begin{figure}[H]
  \centering
  \begin{subfigure}[t]{0.49\textwidth}
    \includegraphics[width=\linewidth]{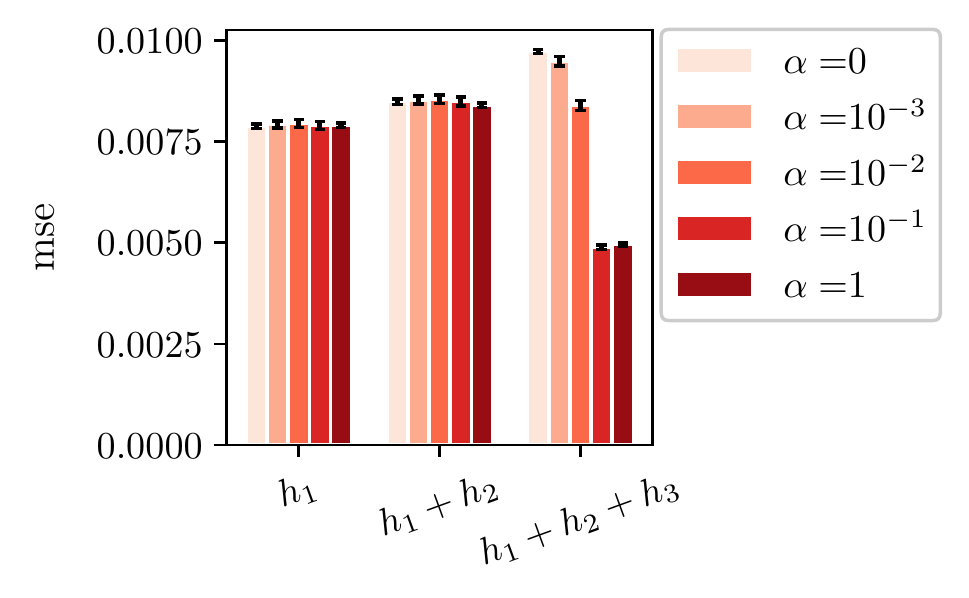}
    \caption{}
  \end{subfigure}
  \begin{subfigure}[t]{0.49\textwidth}
    \includegraphics[width=\linewidth]{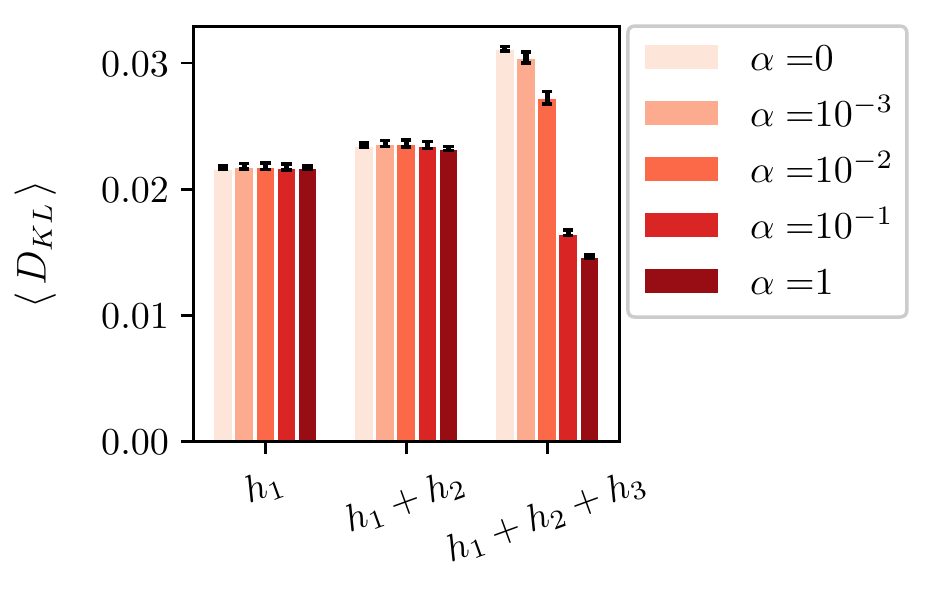}
    \caption{}
  \end{subfigure}
  \caption{\textbf{\textit{logit} approach:} performance measures in terms of (a) mean squared error  and (b) mean Kullback-Leibler divergence. Notations and parameters are same as in fig.~\ref{fig_alpha_clipping}}
\label{fig_alpha_logit}
\end{figure}

On the other hand, change in \textit{ridge parameter} affects the performance in \textit{logit} approach, especially with third order monomials; see fig.~\ref{fig_alpha_logit}. Better performance can be achieved with suitable parameter choice, say $\alpha = 10^{-1}$. 

\pagebreak

\section{Short training data and double descent curve}

\begin{figure}[H]
\includegraphics[scale=1]{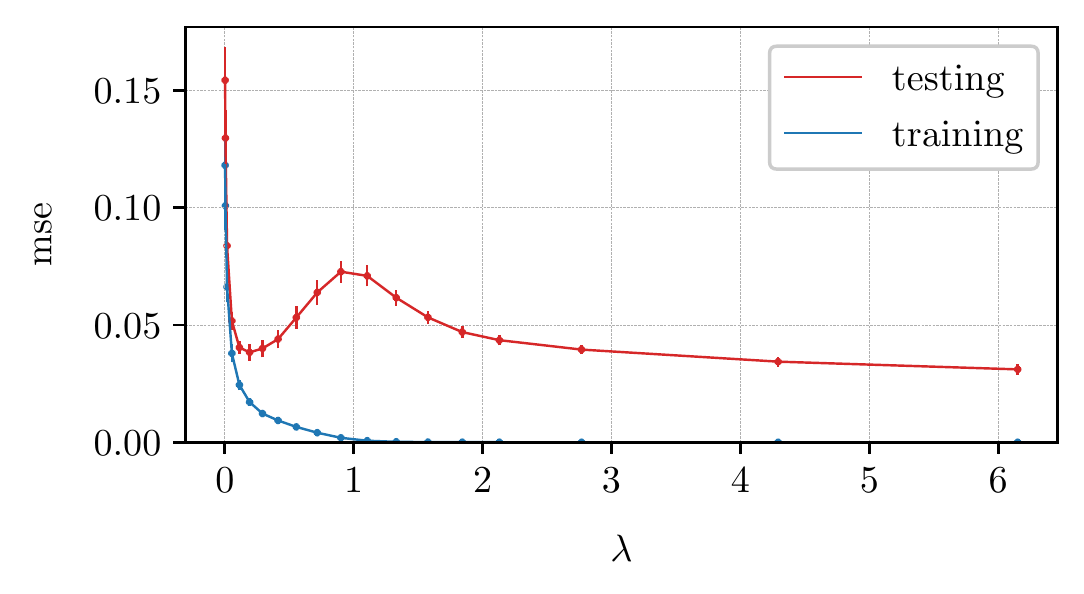}
\caption{
The training error (blue) decreases monotonically as function of the ratio $\lambda$
between the number of fit parameters (weights) and the size of the training data.
In contrast, the testing error (red) displays a local minimum, and attains a peak close to $\lambda=1$, 
before it descents again in the over-fitting regime for $\lambda>1$.
Parameters: 
$\gamma=5\, \mathrm{s}^{-1}$,
$r=2\,\mathrm{s}$, 
$D=0.5\, \mathrm{s}^{-1}$, 
duration of training data $= 6\,\mathrm{s}$, duration of testing data $= 40\,\mathrm{s}$,
$\Delta t= 0.005\,\mathrm{s}$, 
$N=2$,
ridge parameter $\alpha = 10^{-3}$. 
Reported results represent mean$\pm$s.e.m.\ from $5$ realizations.
}
\label{fig_ratio}
\end{figure}

The nVAR machine gives acceptable results even for shorter training data sets. 
However, its performance strongly depends on the number of weights that has to be optimized. 
To highlight this point, we consider the ratio
\begin{equation}
\lambda = \frac{\text{size of weight matrix}}{\text{size of the training data}} \quad.
\end{equation}
Fig.~\ref{fig_ratio} shows the mean squared error (mse) as a function of this ratio $\lambda$. 
Here, the size of the training data was kept fixed, while the delay $k$ was varied. 
Note that the number of weights increases polynomially with $k$, see also fig.~{2 (a)} of the main text. 
The training error monotonically decreases to zero if $\lambda$ increases,
approaching virtually zero in the over-fitting regime with $\lambda>1$.
In contrast, the testing error changes non-monotonically with $\lambda$:
the testing error initially decreases to a minimum, 
then increases and attains a peak close to $\lambda=1$. 
In the over-fitting regime for $\lambda>1$, 
the testing error once again decreases. 
We refer to this non-monotonic depence on $\lambda$ as ``double descent curve'', similar to \cite{loog2020}.

\bibliographystyle{unsrt}
\bibliography{supplement}